# Binary Sequences with Minimum Peak Sidelobe Level up to Length 68

Leukhin A.N., Potehin E.N.

*Povolgskii State Technical University*

+7(8362) 68-78-42, +7(8362) 45-53-73, leukhinan@list.ru, potegor@yandex.ru

Abstract: Results of an exhaustive search for minimum peak sidelobe level binary sequences are presented. Several techniques for efficiency implementation of search algorithm are described. A table of number of non-equivalent optimal binary sequences with minimum peak sidelobe (MPS) level up to length 68 is given. This number can be used in prediction of the longest length for a given sidelobe level of binary sequences. The examples of optimal binary MPS sequences having high merit factor are shown.

*Exhaustive search, binary, minimum peak sidelobe, sequences, aperiodic, autocorrelation function, merit factor, optimal, peak sidelobe*

## Introduction

Binary sequences with low autocorrelation sidelobe levels are useful in different applications: radar, communication systems, information security, synchronization and so on. The main point in radar is to gain the signal-to-noise ratio benefits of a long pulse along the range resolution of a short pulse. This requires longer and longer sequences, with low aperiodic autocorrelation sidelobes. There are no theoretical methods to generate such sequences, so they have been produced by computer searches. The computational complexity of such searches increases with length: a search for N=71 is approximately twice as difficult as that for N=70 that is why computational time doubles.

This paper adds to available knowledge for record length of binary MPS sequence and provides a number of non-equivalence classes for each lengths up to $N = 68$. Despite the computational challenges, progress has been made over time due to improvements in both computational resources and search methods.

Lindner [1] in 1975 did an exhaustive search for binary MPS sequences up to $N = 40$. Cohen et al. [2] in 1990 continued up to $N = 48$. Coxson and Ruso [3] executed an exhaustive search of binary MPS sequences for $N = 64$. So binary MPS sequences are only known for the lengths up to $N = 48$ and for the length $N = 64$.

Apart from known results of global exhaustive search of binary MPS sequences, there are some useful results of local search of binary sequences with the low aperiodic autocorrelation. Kerdock et al. [4] in 1986 found binary sequences for lengths $N = 51, 69, 88$ with $PSL = 3, 4, 5$ respectively. These are still the best. Elders-Boll et al. [5] in 1997 found best known binary PSL sequences with $PSL = 4$ for the lengths from $N = [49; 61]$. Coxson and Ruso [3] in 2004 continued the list of best known binary PSL sequences with $PSL = 4$ up to $N = 70$. Nunn and Coxson [6] in 2008 found best known binary PSL sequences with $PSL = 4$ up to $N = 82$ and with $PSL = 5$ for $N = [83, 105]$.

We have made an exhaustive search of binary MPS sequences during 4 months implementation 1 supercomputer Flagman RX240T8.2 on the base of 8 NVIDIA TESLA C2059 with 3584 parallel graphical processors and on the base of 2 processors Intel Xeon X5670 (up to Six-Core) and using CUDA compilation. Our algorithm is based on a concept of Mertens [7], the branch-and-bound algorithm, using new assembler instructions for calculation of aperiodic autocorrelation function and "package" regime.

## Preliminaries

A binary sequence of length $N$ is an $N$-tuple $A = (a_0, a_1, ..., a_{N-1})$ where each $a_n \in \{-1, 1\}$, $n = 0, 1, ..., N-1$. The aperiodic autocorrelation of $A$ at shift $\tau$ defined as

$$C_\tau = \sum_{n=0}^{N-1-\tau} a_n \cdot a_{n+\tau}. \quad (1)$$

There are two principal measures of level of sidelobe level. The primary measure is the peak sidelobe level (PSL):

$$PSL(C) = \max_{1 \leq \tau \leq N-1} |C_\tau|. \quad (2)$$

For optimal binary sequences by PSL criteria the peak sidelobe has to be minimum:

$$MPS = \min_A PSL. \quad (3)$$

A secondary measure, is the merit factor (MF):

$$MF(C) = \frac{N^2}{2\sum_{\tau=1}^{N-1}[C_\tau]^2}. \qquad (4)$$

PSL affects the maximum of self interference of the sequence and merit factor determine average interference. There are three operations that preserve peak sidelobe level in binary codes: reversal: $R(a_n) = a_{N-1-n}$, negation: $N(a_n) = -a_n$, alternating sign $S(a_n) = (-1)^n a_n$. The sequences obtained within such transformations will be formed class of equivalence.

We are interested to find all non-equivalent classes of binary MPS sequences for each length $N$ from the range $N = [49, 68]$.

# Effective global algorithm for exhaustive search of binary PSL sequences

We modified all achievements of Nunn and Coxson algorithm which based on the main idea of Mertens's branch-and-bound algorithm as described below.

*Base idea of branch-and-bound algorithm.* Let us consider the binary sequence from two opposite sides of the length $N/2$. We called them left and right half-sequences. In the first step we ? choose left bit $a_0$ and right bit $a_{N-1}$. They can be considered like new code $a_0 \, a_{N-1}$. This pair can be "00", "01", "10" and "11". We present the algorithm in graph form, where possible pairs of opposite bits can be shown like leaves of the tree:

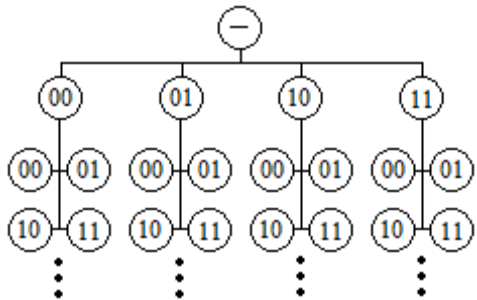

Fig. 1 Set of sequences presented in tree manner

We can check the PSL level of autocorrelation during each step. If on the next step the PSL is not valued all branch of the tree can be excluded from searching. Also due to operations that preserve peak sidelobe level in binary codes we can say that on the first step one pair $a_0 = 0$, $a_{N-1} = 0$ is formed whole class of equivalence for all other possible values $a_0\ a_{N-1}$. So we the only one branch instead of for. On the second? step we have only 3 non-equivalent branches

$$0,0,a_2,...,a_{N-3},0,0, \quad 0,0,a_2,...,a_{N-3},1,0, \quad 0,1,a_2,...,a_{N-3},1,0.$$

The search space can be reduced by excluding equivalent branches, as shown in Fig.2.

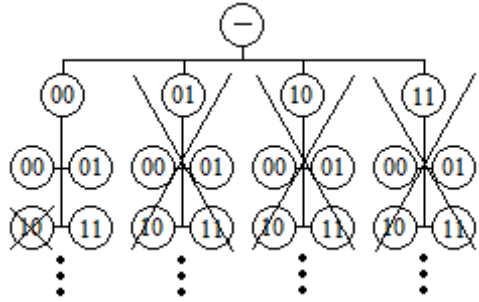

Fig.2. Excluding of the equivalent sequences

*Our modifications.*

1. We used recursive implementation of our algorithm (Fig.3) using inline options for all external operations.

2. Our main idea is to use new assembler instructions for computing autocorrelation function of binary sequences. We can find side lobes of aperiodic autocorrelation using XOR operation. To determine the level of sidelobe we have to calculate the numbers of zeros and units for each shift of sequences. New Intel processors have microarchitecture Intel Core of version SSE4.2 which operating with the set of command on low level. For example C/C++ Microsoft compiler has function __popcnt64 of intrin library and also compilersGCC and G++ has function _mm_popcnt_u64 of smmintrin library for calculation the number of units in binary sequences by 1 cycle.

3. For excluding equivalent sequences we used reverse transformation for two bytes at the time instead of each bit. All possible reverses are stored in static massive with 65536 different bit variations.

4. Also we realized parallel computing in multiprocessor system for all set of non-equivalent sequences separately each from other. We implemented our algorithm on CUDA SDK using function __popcll() for calculating number of unit bits.

5. Finally we used "package" regime to find some binary PSL sequences with the lengths $N, N+2, N+4,...$, because cross correlation functions for left and right parts of the sequences with lengths $N, N+2, N+4,...$, are the same.

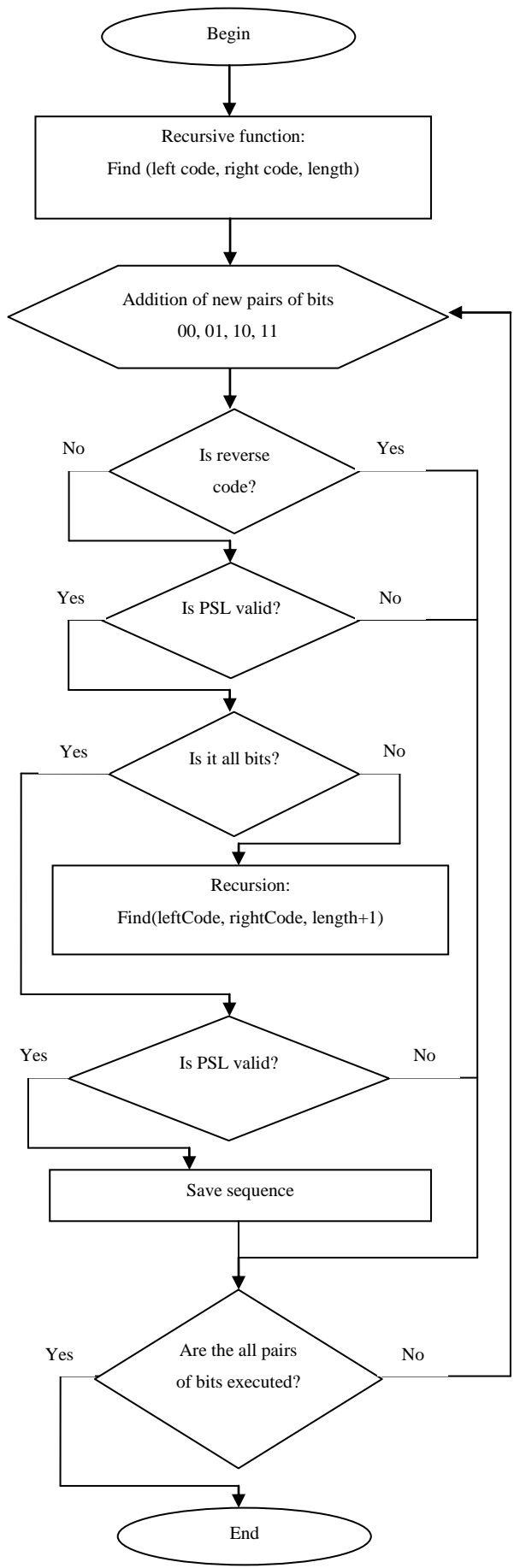

Fig.3 Block-scheme of recursive algorithm

# Results of exhaustive search of binary MPS sequences

Our results are presented in Table 1. There are number of binary PSL sequences, which means that PSLi sequences have exactly $PSL = i$, not less. Synthesized sequences are available on our website [7].

Table 1. Size of set non-equivalent PSL sequences

| Length $N$ | Size of set non-equivalent PSL sequences ||||| 
|---|---|---|---|---|---|
| | PSL1 | PSL2 | PSL3 | PSL4 | PSL5 |
| 2 | 1 | 0 | 0 | 0 | 0 |
| 3 | 1 | 1 | 0 | 0 | 0 |
| 4 | 1 | 1 | 1 | 0 | 0 |
| 5 | 1 | 3 | 1 | 1 | 0 |
| 6 | 0 | 4 | 4 | 1 | 1 |
| 7 | 1 | 7 | 5 | 5 | 1 |
| 8 | 0 | 8 | 12 | 8 | 6 |
| 9 | 0 | 10 | 23 | 20 | 29 |
| 10 | 0 | 5 | 46 | 35 | 30 |
| 11 | 1 | 7 | 53 | 97 | 52 |
| 12 | 0 | 16 | 87 | 133 | 152 |
| 13 | 1 | 11 | 126 | 287 | 246 |
| 14 | 0 | 9 | 152 | 486 | 583 |
| 15 | 0 | 13 | 223 | 800 | 1050 |
| 16 | 0 | 10 | 361 | 1173 | 2176 |
| 17 | 0 | 4 | 307 | 2243 | 3490 |
| 18 | 0 | 2 | 339 | 3025 | 7205 |
| 19 | 0 | 1 | 419 | 4661 | 11645 |
| 20 | 0 | 3 | 625 | 6245 | 21456 |
| 21 | 0 | 3 | 505 | 9826 | 32539 |
| 22 | 0 | 0 | 378 | 11840 | 58331 |
| 23 | 0 | 0 | 515 | 16533 | 86812 |
| 24 | 0 | 0 | 858 | 20673 | 148583 |
| 25 | 0 | 1 | 436 | 29794 | 206762 |
| 26 | 0 | 0 | 242 | 31205 | 329356 |
| 27 | 0 | 0 | 388 | 40193 | 469454 |
| 28 | 0 | 2 | 624 | 49884 | 753204 |
| 29 | 0 | 0 | 284 | 63059 | 966451 |
| 30 | 0 | 0 | 86 | 59506 | 1390617 |
| 31 | 0 | 0 | 251 | 71546 | processing |
| 32 | 0 | 0 | 422 | 89190 | 2894816 |
| 33 | 0 | 0 | 139 | 98644 | processing |
| 34 | 0 | 0 | 51 | 84636 | 4567602 |
| 35 | 0 | 0 | 111 | 98331 | processing |
| 36 | 0 | 0 | 161 | 118624 | 8951507 |
| 37 | 0 | 0 | 55 | 119053 | processing |
| 38 | 0 | 0 | 17 | 89067 | 11788025 |

Table 1. Size of set non-equivalent PSL sequences (continuing)

| | | | | | |
|---|---|---|---|---|---|
| 39 | 0 | 0 | 30 | 101808 | processing |
| 40 | 0 | 0 | 57 | 118731 | 22333659 |
| 41 | 0 | 0 | 15 | 112039 | processing |
| 42 | 0 | 0 | 4 | 72716 | 24453952 |
| 43 | 0 | 0 | 12 | 83417 | processing |
| 44 | 0 | 0 | 15 | 98334 | 44270683 |
| 45 | 0 | 0 | 4 | 82538 | processing |
| 46 | 0 | 0 | 1 | 47331 | 41354620 |
| 47 | 0 | 0 | 1 | 54896 | processing |
| 48 | 0 | 0 | 4 | 64424 | 74010972 |
| 49 | 0 | 0 | 0 | 49088 | processing |
| 50 | 0 | 0 | 0 | 25169 | 57294359 |
| 51 | 0 | 0 | 1 | 28249 | processing |
| 52 | 0 | 0 | 0 | 33058 | processing |
| 53 | 0 | 0 | 0 | 23673 | processing |
| 54 | 0 | 0 | 0 | 10808 | processing |
| 55 | 0 | 0 | 0 | 11987 | processing |
| 56 | 0 | 0 | 0 | 15289 | processing |
| 57 | 0 | 0 | 0 | 9476 | processing |
| 58 | 0 | 0 | 0 | 4026 | processing |
| 59 | 0 | 0 | 0 | 4624 | processing |
| 60 | 0 | 0 | 0 | 5542 | processing |
| 61 | 0 | 0 | 0 | 3246 | processing |
| 62 | 0 | 0 | 0 | 1212 | processing |
| 63 | 0 | 0 | 0 | 1422 | processing |
| 64 | 0 | 0 | 0 | 1859 | processing |
| 65 | 0 | 0 | 0 | 1003 | processing |
| 66 | 0 | 0 | 0 | 324 | processing |
| 67 | 0 | 0 | 0 | 414 | processing |
| 68 | 0 | 0 | 0 | 491 | processing |

The results of an exhaustive search of binary MPS sequences up to length $N = 68$ are presented in Table 2. Also in the Table 2 there are the highest level of MF between binary MPS sequences and examples of such sequences in hexadecimal format.

Table 2. Results of exhaustive search of binary MPS sequences

| Length | PSL | MF | Optimal or best known by MF? | Sequence | Size of set |
|---|---|---|---|---|---|
| 2 | 1 | 2 | yes | 0 | 1 |
| 3 | 1 | 4,5 | yes | 1 | 1 |
| 4 | 1 | 4 | yes | 2 | 1 |

Table 2. Results of exhaustive search of binary MPS sequences (continuing)

| | | | | | |
|---|---|---|---|---|---|
| 5 | 1 | 6,25 | yes | 02 | 1 |
| 6 | 2 | 2,571 | yes | 02 | 4 |
| 7 | 1 | 8,167 | yes | 0D | 1 |
| 8 | 2 | 4 | yes | 1A | 8 |
| 9 | 2 | 3,375 | yes | 02C | 10 |
| 10 | 2 | 3,846 | yes | 02C | 5 |
| 11 | 1 | 12,1 | yes | 0ED | 1 |
| 12 | 2 | 7,2 | yes | 0A6 | 16 |
| 13 | 1 | 14,083 | yes | 00CA | 1 |
| 14 | 2 | 5,158 | yes | 00CA | 9 |
| 15 | 2 | 4,891 | no | 0329 | 13 |
| 16 | 2 | 4,571 | no | 1DDA | 10 |
| 17 | 2 | 4,516 | yes | 0192B | 4 |
| 18 | 2 | 6,48 | yes | 0168C | 2 |
| 19 | 2 | 4,878 | no | 0EEDA | 1 |
| 20 | 2 | 5,263 | no | 04D4E | 3 |
| 21 | 2 | 6,485 | no | 005D39 | 3 |
| 22 | 3 | 6,205 | yes | 013538 | 378 |
| 23 | 3 | 5,628 | yes | 084BA3 | 515 |
| 24 | 3 | 8 | yes | 31FAB6 | 858 |
| 25 | 2 | 7,102 | no | 031FAB6 | 1 |
| 26 | 3 | 7,511 | no | 07015B2 | 242 |
| 27 | 3 | 9,851 | yes | 0F1112D | 388 |
| 28 | 2 | 7,84 | yes | 4B7770E | 2 |
| 29 | 3 | 6,782 | yes | 04B7770E | 284 |
| 30 | 3 | 7,627 | yes | 03F6D5CE | 86 |
| 31 | 3 | 7,172 | yes | 07E736D5 | 251 |
| 32 | 3 | 7,111 | no | 01E5AACC | 422 |
| 33 | 3 | 8,508 | yes | 003CB5599 | 139 |
| 34 | 3 | 8,892 | yes | 0CC01E5AA | 51 |
| 35 | 3 | 7,562 | no | 00796AB33 | 111 |
| 36 | 3 | 6,894 | no | 3314A083E | 161 |
| 37 | 3 | 6,985 | no | 031D5AD93F | 55 |
| 38 | 3 | 8,299 | yes | 003C34AA66 | 17 |
| 39 | 3 | 6,391 | no | 13350BEF3C | 30 |
| 40 | 3 | 7,407 | yes | 2223DC3A5A | 57 |
| 41 | 3 | 7,504 | no | 038EA520364 | 15 |
| 42 | 3 | 8,733 | yes | 04447B874B4 | 4 |

Table 2. Results of exhaustive search of binary MPS sequences (continuing)

| | | | | | |
|---|---|---|---|---|---|
| 43 | 3 | 6,748 | no | 005B2ACCE1C | 12 |
| 44 | 3 | 6,286 | no | 202E2714B96 | 15 |
| 45 | 3 | 6,575 | no | 02AF0CC6DBF6 | 4 |
| 46 | 3 | 6,491 | no | 03C0CF7B6556 | 1 |
| 47 | 3 | 7,126 | no | 069A7E851988 | 1 |
| 48 | 3 | 6,128 | no | 24AC8847B87C | 4 |
| 49 | 4 | 8,827 | yes | 05E859E984451 | 49088 |
| 50 | 4 | 8,17 | yes | 038FE23225492 | 25169 |
| 51 | 3 | 7,517 | no | 0E3F88C89524B | 1 |
| 52 | 4 | 8,145 | yes | 05FB6D5D9D8E3 | 33058 |
| 53 | 4 | 7,89 | no | 00FF66EAE96B1C | 23673 |
| 54 | 4 | 7,327 | no | 043B48A28793B3 | 10808 |
| 55 | 4 | 7,451 | no | 1658A2BC0A133B | 11987 |
| 56 | 4 | 8,167 | yes | 0C790164F6752A | 15289 |
| 57 | 4 | 7,963 | no | 01B4DE3455B93BF | 9476 |
| 58 | 4 | 8,538 | yes | 008D89574E1349E | 4026 |
| 59 | 4 | 8,328 | no | 1CAD63EFF126A2E | 4624 |
| 60 | 4 | 8,108 | no | 119D01522ED3C34 | 5542 |
| 61 | 4 | 7,563 | no | 0024BA568EB83731 | 3246 |
| 62 | 4 | 8,179 | yes | 000C67247C59568B | 1212 |
| 63 | 4 | 9,587 | yes | 1B3412F0501539CE | 1422 |
| 64 | 4 | 9,846 | yes | 26C9FD5F5A1D798C | 1859 |
| 65 | 4 | 8,252 | no | 04015762C784EC369 | 1003 |
| 66 | 4 | 7.751 | no | 03FEF2CCB0B8CAC54 | 324 |
| 67 | 4 | 7.766 | no | 073C2FADC44255264 | 414 |
| 68 | 4 | 8.438 | no | 562B8CA48E0C9027E | 491 |

## Conclusion

Optimal binary MPS sequences are updated for lengths 49 to 63 and 65 to 68. Also the number of non-equivalent PSL1, PSL2, PSL3, SPL4 and PSL5 sequences are found for lengths 2 to 68. The number of MPS and PSL sequences are useful for prediction of longest sequences possible for a given PSL. Such optimal sequences are highly sought after in radar and cryptography. These sequences were considered from MF criteria and the MPS sequences with highest MF were identified. The number of PSL is sequences rapidly increased for increasing PSL. For example for the length 48 the number of nonequivalent binary sequences with PSL3 is 4, PSL4 is 64424, PSL5 is 74010972.

Acknowledgements. This work is supported by the grants: grant RFBR №12-07-00552, project №1.07.2012.

Preparing for

INTERNATIONAL WORKSHOP ON CODING AND CRYPTOGRAPHY, WCC 2013, April 15-19, 2013, Bergen, Norway.

Date: 14 December 2012.